\documentclass[aps, pra, amsmath, showpacs, preprintnumbers, twocolumn, amssymb]{revtex4}

\usepackage{graphicx}
\usepackage{mathptmx, textcomp}
\usepackage[latin1]{inputenc}
\usepackage{tabularx}
\usepackage{multirow}
\usepackage{color}
\usepackage{pifont}
\definecolor{dgreen}{RGB}{46,139,87}
\begin{document}

\title{Resonant atom-dimer collisions in cesium: Testing universality at positive scattering lengths}

\author{A. Zenesini$^{1,2}$, B. Huang$^1$, M. Berninger$^1$, H.-C. N\"{a}gerl$^1$, F. Ferlaino$^1$, R. Grimm$^{1,3}$}

\address{$^1$ Institut f\"ur Experimentalphysik and Zentrum f\"ur Quantenphysik, Universit\"at Innsbruck, 6020 Innsbruck, Austria}
\address{$^2$ Institute of Quantum Optics, Leibniz Universit\"at Hannover, Welfengarten 1, 30167 Hannover, Germany}
\address{$^3$ Institut f\"ur Quantenoptik und Quanteninformation, \"Osterreichische Akademie der Wissenschaften, 6020 Innsbruck, Austria}

\date{\today}

\begin{abstract}
We study the collisional properties of an ultracold mixture of cesium atoms and dimers close to a Feshbach resonance near 550\,G in the regime of positive $s$-wave scattering lengths. We observe an atom-dimer loss resonance that is related to Efimov's scenario of trimer states. The resonance is found at a value of the scattering length that is different from a previous observation at low magnetic fields. This indicates non-universal behavior of the Efimov spectrum for positive scattering lengths. We compare our observations with predictions from effective field theory and with a recent model based on the van der Waals interaction. We present additional measurements on pure atomic samples in order to check for the presence of a resonant loss feature related to an avalanche effect as suggested by observations in other atomic species. We could not confirm the presence of such a feature.
\end{abstract}

\maketitle


\section{Introduction}

Efimov's solution to the problem of three resonantly interacting particles \cite{Efimov1970ela} is widely considered to be the most prominent example of a \textit{universal} few-body system, where the knowledge of the two-body scattering length $a$ and an additional three-body parameter is sufficient to define the whole energy spectrum and to locate all the bound states. The details of the interparticle potential become irrelevant and different systems very far apart in energy and length scales can be described in the same way. The famous discrete scaling of the Efimov spectrum (scaling factor of $22.7$) and the precise ratios that link its different parts have attracted large interest in the scientific community.

Universal behavior arises from the presence of resonant interactions leading to collisions on a length scale exceeding the typical size of the interparticle potential.  In trimer systems, the contributions of the short-range details are commonly included in the ``three-body parameter". This parameter fixes the starting point of the spectrum and can be expressed in terms of the scattering length $a_-$ at which the most deeply bound Efimov state crosses the zero-energy threshold \cite{Braaten2006uif}. Within the ideal Efimov scenario, the positions of all the other features of the spectrum are uniquely determined, both at positive and negative values of $a$.

The validation of Efimov's scenario had remained elusive for decades until experiments on ultracold atoms provided evidence for its existence \cite{Kraemer2006efe, Knoop2009ooa, Zaccanti2009ooa, Gross2009oou, Gross2010nsi, Pollack2009uit, Wild2012mot, Berninger2011uot, Huang2014oot, Tung2014oog, Pires2014ooe}. The appearance of trimer bound states has been shown by measuring inelastic collisional rates in atomic samples or atom-dimer mixtures by tuning the scattering length via magnetically controlled Feshbach resonances \cite{Chin2010fri}. The presence of trimer bound states leads to enhancement and suppression of losses \cite{Esry1999rot, Braaten2001tbr, Petrov2004tbp}. In particular the loss resonances represent a ``smoking gun" for Efimov's spectrum and occur where the trimer energy state crosses the atomic threshold (at $a_-$, in the region of negative $a$) or merges into the state of a dimer plus a free atom (at $a_*$, in the region of positive $a$).

In the region of negative scattering lengths, experimental observations have shown that the position $a_-$ is essentially independent of the particular Feshbach resonance used for interaction tuning \cite{Berninger2011uot, Gross2009oou, Roy2013tot}. The comparison between experiments performed with different species \cite{Zaccanti2009ooa, Gross2009oou, Gross2010nsi, Pollack2009uit, Wild2012mot} shows that  $a_-\approx-9.5\,R_{\rm vdW}$, where the van der Waals length $R_{\rm vdW}$ represents the length scale associated with the van der Waals interaction \cite{Chin2010fri}. This result suggested that the knowledge of $R_{\rm vdW}$ is sufficient to determine the three-body parameter. This idea is supported by theoretical results for the region of negative scattering lengths \cite{Wang2012oft, Schmidt2012epb, Chin2011uso, Naidon2012poo, Naidon2014moa}, pointing to a new type of universality, named ``van der Waals universality", in atomic systems.

In the region of positive scattering lengths, the most suitable observables are atom-dimer resonances, as detected by enhanced losses in mixtures of atoms and dimers \cite{Knoop2009ooa, Bloom2013tou,Lompe2010ads, Huckans2009tbr, Nakajima2010nea, Nakajima2011moa}. They provide more direct and unambiguous evidence in contrast to related recombination minima and avalanches in atomic samples \cite{Zaccanti2009ooa, Gross2009oou, Pollack2009uit}. The essential prerequisite for studying inelastic atom-dimer collisions is the existence of efficient methods to convert atoms into dimers in a controlled manner. First measurements on atom-dimer mixtures were performed in samples consisting of Cs atoms and magneto-associated Cs dimers \cite{Knoop2009ooa}. Cesium represents an ideal candidate because of the rich Feshbach spectrum and the good atom-dimer conversion efficiency \cite{Mark2007sou}. These measurements, which were performed in the low magnetic field region, gave a first hint on large deviations from universal ratios \cite{Knoop2009ooa}. This result was unexpected as the experiments performed with atomic samples of Cs, in the same region of scattering lengths, have shown recombination minima at scattering length values consistent with universal behavior \cite{Kraemer2006efe}. Other experimental investigations have been performed with atom-dimer mixtures of different $^6$Li hyperfine sublevels \cite{Lompe2010ads, Huckans2009tbr, Nakajima2010nea, Nakajima2011moa} and in heteronuclear mixtures of fermionic $^{40}$K$^{87}$Rb dimers and $^{87}$Rb atoms \cite{Bloom2013tou}.

Theoretical models qualitatively explain the deviations in the relative positions of the atom-dimer features of the spectrum \cite{Hammer2007erc, Thogersen2008upo, Dincao2009tsr, Platter2009rct} by introducing finite-range effects. However, the predicted corrections are too small to explain the observed resonance positions. Recently, theoretical work \cite{Wang2013uvd} has proposed a simple two-spin model to directly include the van der Waals (vdW) interaction into the Efimov problem for atomic systems. With this model, which we will refer to as ``vdW model", the predicted position of the atom-dimer resonance in Cs is in good agreement with the experimental observation made in the region of low magnetic field. The authors were also able to explain the deviation from the ideal scaling relation between positions of the triatomic and the atom-dimer loss resonances.

In this Article, we explore inelastic atom-dimer scattering in an ultracold mixture of Cs atoms and Cs$_2$ Feshbach molecules near a broad $s$-wave resonance located at about 550\,G \cite{Berninger2013frw}. Our measurements on the atom-dimer collisional rate coefficient $\beta$ reveal a pronounced resonant feature similar to the one observed at low magnetic fields in our previous work \cite{Knoop2009ooa}. Our findings reveal a significant difference in the positions of the two atom-dimer resonances. This suggests that universality is much less robust for positive values of $a$ as compared to the negative-$a$ region. Nevertheless, we show that our observations can be quantitatively predicted in the framework of the universal vdW model according to Ref.\,\cite{Wang2013uvd}.

This Article is organized as follows. In Section \ref{Procedure} we describe the experimental procedure to create ultracold samples of atoms and molecules. In Section \ref{Decay}, we explain the measurement of the atom-dimer decay rates. Section \ref{discussion} contains the comparison with previous results and available theoretical models. In Section \ref{avalanche} we discuss our experimental search for avalanche loss processes in samples of Cs atoms. In Section \ref{conclusion} we present our conclusions.

\begin{figure}[b]
\begin{centering}
\includegraphics[width=.49\textwidth] {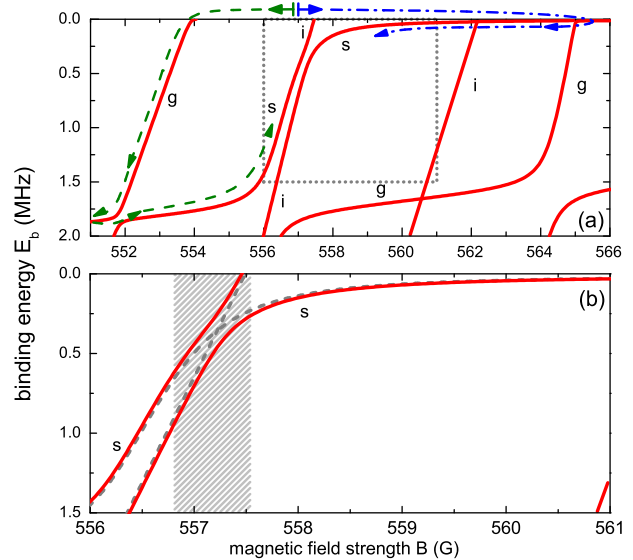}
\caption{(color online) Near-threshold energy spectrum of Cs$_2$ in the magnetic-field region around 560\,G. (a) The spectrum results from three $g$-, two $i$-, and two $s$-wave molecular states. The bent $s$-wave state is our target state for atom-dimer decay measurements and the $g$-wave states are used to prepare molecules in the $s$-wave state. The paths for molecule creation are indicated by the dashed and dash-dotted arrows. The dotted frame indicates the region of interest as magnified in panel (b). (b) The large $i/s$ avoided crossing is highlighted by the difference between the uncoupled states ($s$, $d$, $g$ basis set and $i$-wave state \cite{Berninger2013frw}), shown as dashed lines, and the coupled states (solid lines); for details see text. The hatched region marks the range of $B$ for which the $s$-wave character of the states is below 90\%.}
\label{energy}
\end{centering}
\end{figure}

\section{Experimental procedures}
\label{Procedure}

Our experiments are performed with an ultracold sample of Cs atoms in the ground-state sublevel $|F=3, m_F=3>$, where $F$ is the hyperfine and $m_F$ the magnetic quantum number. First we prepare the sample at high magnetic fields following the procedure described in Ref.\,\cite{Berninger2013frw}. We then convert a fraction of the atoms into Feshbach molecules by magneto-association \cite{Chin2010fri, Herbig2003poa}. More details on the near-threshold molecular structure, including the relevant quantum numbers, can be found in Ref. \,\cite{Berninger2013frw}.

The atomic sample is evaporatively cooled in a crossed optical dipole trap generated by near-infrared single-frequency laser light at a wavelength of 1064.5\,nm. As discussed in Ref.\,\cite{Berninger2013frw} we take advantage of a broad open-channel dominated $s$-wave Feshbach resonance to control the elastic collisional rate during the evaporation stage. This resonance is centered at a magnetic field of $B=548.78(9)$\,G and has a width of 7.5(1)\,G \cite{note1}. Different from the experimental procedure described in Ref\,\cite{Berninger2013frw}, in the last part of the evaporation stage we set $B$ to $556.9$\,G, corresponding to a scattering length $a$ of about $280\,a_0$, where $a_0$ is Bohr's radius \cite{note2}. We typically obtain $1.5\times 10^5$ atoms at a temperature $T\approx$150\,nK. The final trap has a mean frequency of $\bar{\omega} = 2\pi \times 27.1(2)$\,Hz. This non-condensed sample with a peak number density of $1.6 \times 10^{12}$ cm$^{-3}$ and a peak phase-space density of about 0.1 is our starting point for the creation of dimers.

Cesium exhibits a rich collection of Feshbach resonances \cite{Berninger2013frw, chin2004pfs}. These can be used for magneto-association of atoms to molecules both in the low and in the high magnetic field regions. The near-threshold energy spectrum of the different molecular states in the magnetic-field region of interest is shown in Fig.\,\ref{energy}. In absence of any coupling between the different energy states, the spectrum of the bare molecular states would show essentially straight lines: two states, an $s$- and a $g$-wave state, nearly parallel to the threshold ($E_b=0$) with binding energies of about 25\,kHz and 1.75\,MHz \cite{note3}, respectively, and five other states (one $s$-, two $g$-, and two $i$-wave states) with slopes of about 1 MHz/G relative to threshold. Various coupling mechanisms \cite{Hutson2008acb} lead to a manifold of avoided crossings in the spectrum.

The bent $s$-wave state is the target state for our atom-dimer decay measurements. This state undergoes an avoided crossing with an $i$-wave state at $B=557.25$\,G and at a corresponding binding energy of 400\,kHz. The coupling strength, \textit{i.e.}\,half the energy splitting at the center of the crossing, is about 100\,kHz. This is unusually strong for a crossing between states that differ by six units of angular momentum. The particular mechanism leading to this higher-order crossing is not understood in the framework of the available theory \cite{Berninger2013frw}. The coupled-channel model presented in Ref.\,\cite{Berninger2013frw} can accurately determine the positions and the coupling strengths of states with rotational quantum numbers up to $\ell=4$ in a basis set of $s$, $d$, and $g$ states. We have calculated the energies near the $i/s$-crossing as shown in Fig.\,\ref{energy} by fitting a simple two-level model to the experimental data \,\cite{Berninger2013frw}. One level represents the bent molecular state in the $s$, $d$, $g$ basis, and the other one is the bare $i$-wave state. In Fig.\,\ref{energy}(b) we highlight with hatched shading the range of $B$ in which the mixing reduces the $s$-wave character of the relevant state to below 90\%. We can expect that the interaction physics of the dimers is open-channel dominated only outside of this region.

Because of the strong $i/s$ avoided crossing we populate the $s$-wave state along two different paths similar to Ref.\,\cite{Mark2007sou}: To create $s$-wave molecules on the lower side of the avoided crossing we use the $g$-wave resonance at $B=554.06(2)$\,G; see dashed lines in Fig.\,\ref{energy}. After the magneto-association the magnetic field $B$ is slowly ramped down to 551\,G to populate the $g$-wave state with binding energy of about $2$\,MHz by adiabatically following the $g/g$-wave avoided crossing near $552$\,G. A subsequent fast upward ramp for $B$ allows diabatic transfer through the $g/g$-wave avoided crossing and then to easily access the lower region of the $s$-wave state by adiabatically following the $s/g$-wave crossing. The creation of molecules on the upper side of the $i/s$-wave crossing is achieved through the $g$-wave resonance at 565.48(2)\,G; see dash-dotted path in Fig\,\ref{energy}. Here, the creation of $s$-wave molecules is facilitated by the relatively large coupling between the $s$-wave and the $g$-wave state at 565\,G that allows us to switch from the $g$-wave to the $s$-wave state as we lower $B$. The coupling between the $s$-wave state and a second $i$-wave state at about 562\,G is negligible and that crossing is always followed diabatically. In both cases, we convert about 8\% of the initial atoms into molecules. The final samples contain about $10^5$ atoms and $4\times 10^3$ molecules in thermal equilibrium at a temperature $T\approx$175\,nK, atomic peak density of $9(1)\times 10^{11}\,\rm{cm}^{-3}$, and a molecular peak density of $9.8(2.1)\times 10^{10}\,\rm{cm}^{-3}$.

To determine the number of atoms and the number molecules we first release the mixture from the trap. We use the Stern-Gerlach technique by applying a strong magnetic field gradient for 3\,ms to separate the molecules from the atoms. We then convert the molecules back to atoms by Feshbach dissociation ramps. Molecules above the $s/i$-crossing are dissociated by reversing the association path. For molecules below the $s/i$-crossing we ramp up the magnetic field to dissociate them to atoms via the $i$-wave state. We detect the atoms by standard absorption imaging.

\section{Measurements of atom-dimer decay}
\label{Decay}

Inelastic atom-dimer collisions are quantitatively described by the corresponding rate coefficient $\beta$. We measure this quantity by observing the decay of the number of molecules in the mixture. For this purpose, we record the time evolution of the atom number $N_A$ and the molecule number $N_D$ for different values of $B$, similarly to Ref.\,\cite{Knoop2009ooa}. We carry out additional measurements in pure dimer samples to determine the background losses caused by inelastic dimer-dimer collisions.

The decay of $N_D$ in the trap can be modelled by the rate equation \cite{Knoop2009ooa}
\begin{equation}
\frac {\dot{N}_D}{N_D} = -\beta' \frac{N_A}{V} - \alpha \frac{N_D}{V},
\label{RateEq}
\end{equation}
with an effective volume $V=[2\pi k_BT/(m\bar{\omega}^2)]^{3/2}$. Here, $m$ is the Cs atomic mass. For the loss rate coefficient we use $\beta' = \sqrt{8/27}\times \beta$, where the factor $\sqrt{8/27}$ takes into account the overlap between the atomic and molecular clouds \cite{Knoop2009ooa}. The first term in Eq.~(\ref{RateEq}) accounts for atom-dimer losses, while the second term models dimer-dimer background losses characterized by the rate coefficient $\alpha$. In our samples, the number of atoms is typically 25 times larger than the number of molecules and thus $N_{A}$ can be considered as approximately constant. Three-body losses take place on a timescale much larger than the timescale for atom-dimer losses and are negligible \cite{Kraemer2006efe, Knoop2009ooa}.

\begin{figure}[t]
\begin{centering}
\includegraphics[width=.45\textwidth] {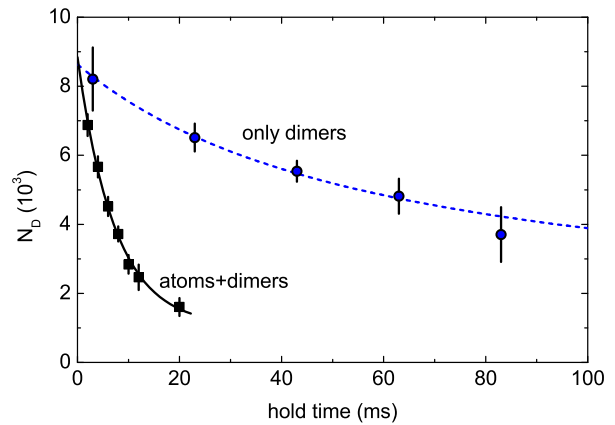}
\caption{(color online) Atom-dimer versus dimer-dimer loss. The number $N_D$ of remaining dimers after a variable hold time for $B=557.7$\,G ($a \approx 400\,a_0$) for a pure molecular sample (circles) and for an atom-dimer mixture (squares). The dashed line is the result of the fit according to a two-body decay rate equation $\dot{N}_D=-\alpha N_D^2/V$ giving $\alpha=2.4(3)\times10^{-10}\rm{cm^3/s}$. The solid line is a fit according to Eq.~(\ref{ND}). The error bars represent the standard deviation for $N_D$ given 5 to 10 experimental runs. Note that, in this particular set of measurements, the initial number of dimers is twice higher than under usual experimental conditions, which enhances dimer-dimer losses.}
\label{alpha}
\end{centering}
\end{figure}

To determine the background contribution of dimer-dimer losses to the measured decay curves, we carry out measurements in pure molecular samples to extract the rate coefficient $\alpha$. For such measurements, we remove the atoms with a pulse of resonant light \cite{Mark2007sou}. An example of a decay measurements on pure molecular samples is shown in Fig.\,\ref{alpha} together with the data obtained with an atom-dimer mixture at the same value for $B$. The larger lifetime of the pure molecular sample is evident and clearly demonstrates that losses in our mixture are dominated by inelastic atom-dimer collisions.

\begin{figure}[t]
\begin{centering}
\includegraphics[width=.45\textwidth] {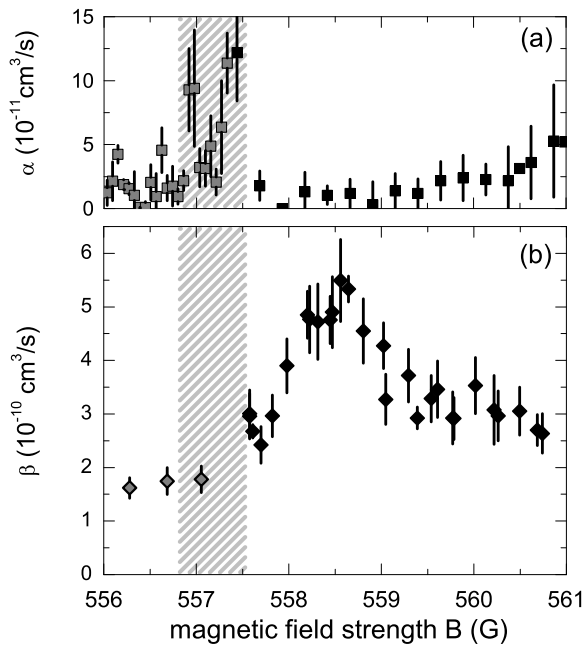}
\caption{Measured dimer-dimer and atom-dimer loss rate coefficients. (a) The dimer-dimer loss rate coefficient $\alpha$ and (b) the atom-dimer loss rate coefficient $\beta$ are plotted as a function of $B$. They are determined from measurements as shown in Fig.\,\ref{alpha}.  The data in (a) is obtained for hold times of up to 10\,ms at the lower side of the avoided crossing (gray squares) and up to 20\,ms at the upper side (filled squares). In (b), gray diamonds are data obtained on the lower side of the avoided crossing and filled diamonds are data on the upper side. The error bars contain the statistical uncertainties on the number of atoms and dimers, trap frequencies, and temperature. As in Fig.\,\ref{energy}, the hatched region indicates the range of $B$ in which the $s$-wave character of the states is below 90\%.}
\label{ADold}
\end{centering}
\end{figure}

Figure \ref{ADold}(a) shows the values of the loss rate coefficient $\alpha$ measured on the two sides of the avoided crossing. In the magnetic field region of interest the coefficient $\alpha$ shows a strong enhancement close to the avoided crossing. We  attribute this to the strong closed-channel contribution in this region. Above the crossing, we observe a behavior resembling previous observations for dimer-dimer collisions in the low-field region \cite{Ferlaino2008cbt}, showing an increase of $\alpha$ for higher magnetic fields where the scattering length $a$ becomes larger. We note that we have also observed two narrow loss features in dimer-dimer collisions at about 556.65(5)\,G and 556.94(5)\,G similar to observations reported in Refs.\,\cite{Chin2005oof, Ferlaino2010cou}. We attribute these features to Feshbach-like resonances, most likely resulting from the coupling of two colliding dimers to a tetramer state.

The atom-dimer relaxation rate coefficient $\beta'$ can now be determined by fitting the molecule number with the solution of Eq.~(\ref{RateEq}) for constant $N_A$,
\begin{equation}
N_D(t)=\frac{\beta' N_AN_{D,0}}{(\beta' N_A+ \alpha N_{D,0})e^{\beta' N_At/V}-\alpha N_{D,0}},
\label{ND}
\end{equation}
where the free parameters are the initial number of molecules $N_{D,0}$ and the rate coefficient $\beta'$, whereas $\alpha$, $N_A$ and $V$ are separately measured quantities. The values obtained for the rate coefficient $\beta$ are displayed in Fig.\ref{ADold}(b) as a function of $B$. It shows a distinct maximum near $B=558.5$\,G in a range of $B$ where the $s$-wave character is dominant and where dimer-dimer losses are very weak. We interpret this feature as an atom-dimer resonance caused by the coupling to an Efimov-like three-body state, in analogy with our previous low-field observation \cite{Knoop2009ooa}.

\section{Comparison with previous results and theory}
\label{discussion}

The observation of an atom-dimer resonance in the region of high magnetic fields as reported here can be compared to the previous observation of an atom-dimer resonance at low magnetic fields \cite{Knoop2009ooa}. This comparison provides a test of the universality of the three-body system for $a>0$ and thus complements our previous work on triatomic Efimov resonances for $a<0$  \cite{Berninger2011uot}.

\begin{figure}[b]
\begin{centering}
\includegraphics[width=.4\textwidth] {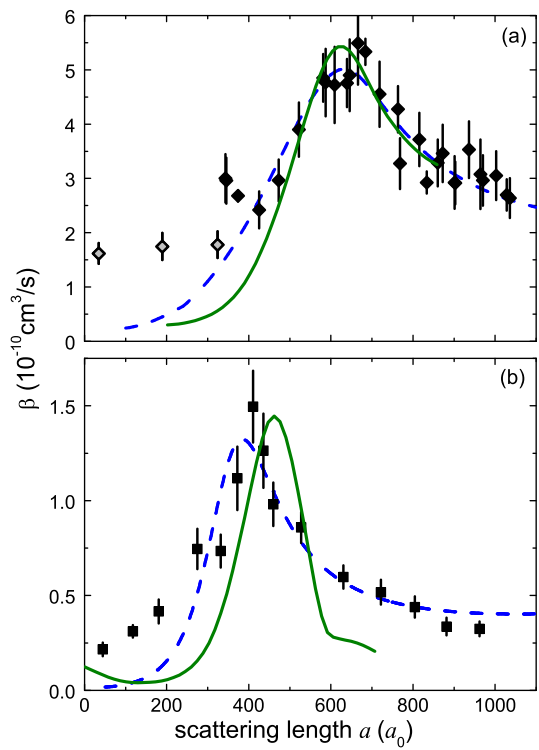}
\caption{(color online) Atom-dimer loss rate coefficient $\beta$ for Cs as a function of the scattering length $a$ in the high (a) and low (b) magnetic field regions. In panel (a) the filled (gray) diamonds result from the measurements performed on the upper (lower) side of the avoided crossing at 175\,nK, as in Fig.\,\ref{ADold}(b). In panel (b) the filled squares are data from Ref\,\cite{Knoop2009ooa} acquired at 170\,nK. In both panels, the dashed and the solid lines represent the EFT fit and the prediction of vdW model scaled by the factors $D$ and $D'$ (see text), respectively. Error bars include statistical uncertainties on temperature, trap frequencies, and atom numbers and the fitting uncertainties.}
\label{comparison}
\end{centering}
\end{figure}

Figure \ref{comparison} presents the atom-dimer loss rate coefficient $\beta$ as a function of the scattering length $a$. Panel (a) shows the new data and panel (b) shows the data from Ref.\,\cite{Knoop2009ooa}. For the $a(B)$ conversion in both data sets, we have used the most recent and very accurate model M2012 \cite{Berninger2013frw}.
The different positions of the two resonant features are evident, with the one observed at high magnetic fields centered at $a\approx 600\,a_0$ and the one at low fields centered at about $400\,a_0$. This difference stands in contrast to the recent observations on triatomic Efimov resonances at negative $a$ \cite{Berninger2011uot}, where the resonance positions appear at essentially the same values of $a$.

In Sec.\,\ref{EFT}, we first present a fit of our experimental data based on effective field theory (EFT), which allows us to extract the parameters describing the atom-dimer resonances. In Sec.\,\ref{waals}, we then compare our observations with a recently developed model \cite{Wang2013uvd} that takes the vdW interaction into account. Finally, in Sec.~\ref{Disc} we, discuss our findings in view of the universality of three-body physics in real atomic systems.

\begin{table*}[ht]
\begin{tabular}{c || c c c || c | c c c c c | c | c c c}
\multirow{2}{*}{$B_{\rm res}$}   	& \multirow{2}{*}{$s_{\rm res}$} 	& \multirow{2}{*}{$a_{\rm bg}/a_0$}
& \multirow{2}{*}{$k_0/a_0$}   & \multirow{2}{*}{$a_-/a_0$ \cite{Berninger2011uot}} 	 &\multicolumn{5}{c|}{EFT fit}& vdW
& \multicolumn{3}{c}{$a_*/|a_-|$} \\
\cline{6-14}
& & & & & $a_*/a_0$ & $\delta_1/a_0$ & $\delta_2/a_0$ & $\eta_*$ & $D$ & $D'$ & Exp & Efim. & vdW\\
\hline
~-12.3\,G~	  			& ~560~    				& ~$\approx$1700~
& ~$180(20)$~				& ~$-872(22)$~				& ~+419(10)~ 			&~8~ & ~6~	
& ~$0.06(2)$~ 				& ~$0.64(23)$~ 			&$0.48(7)$~ 			& ~$0.48(2)$~
& ~$1.06$~  				& ~$0.54$~       			\\

548.8\,G	  				& 170    				& $\approx$2500
& $210(20)$ 				& $-957(80)$   			& $+653(25)$ 				&12 & 22		
& $0.07(2)	$	 			& $2.8(9)$			& $5.6(3)$				& $0.68(6)$	
& $	1.06$ 				& $0.65$          		\\
\end{tabular}
\caption{Parameters for the two Feshbach resonances and the associated atom-dimer resonances. The first column gives the magnetic field value $B_{\rm res}$ for the center of the $s$-wave Feshbach resonance. The quantities $s_{\rm res}$, $a_{\rm bg}$, and $k_0$ are the resonance strength, the background scattering length, and the effective range, respectively \cite{Chin2010fri}. The effective range $k_0$ at the atom-dimer resonance position has been calculated by using the latest Cs potentials. The values for the triatomic resonance positions $a_-$ are taken from Ref.\,\cite{Berninger2011uot} and the errors include all statistical uncertainties. For the values of $a_*$ the number in parentheses gives the full statistical uncertainty, while $\delta_1/a_0$ and $\delta_2/a_0$ are the uncertainties resulting from the fit and from the conversion $a(B)$ \cite{Berninger2013frw}, respectively. The scaling factors $D$ and $D'$ result from the EFT fit \cite{Braaten2001tbr} and from the amplitude fit according to the results from the vdW model. The last three columns give the values for $a_*/|a_-|$ as determined by the experiment, as given by Efimov's universal solution \cite{gogolin2008aso}, and as given by the vdW model. The error for $a_*/|a_-|$ includes all statistical uncertainties.}
\label{fit}
\end{table*}

\subsection{Effective field theory}
\label{EFT}

We analyze our measurements of the loss rate coefficient $\beta$ by using the results of EFT \cite{Braaten2007rdr, Braaten2009erd}. EFT provides a general description of the functional dependence $\beta(a)$, without being able to predict the resonance position and its width. The theory thus contains two free parameters, $a_*$ and $\eta_*$, which are determined by fits to the experimental data. Our fit also includes an additional amplitude scaling factor $D$ to account for systematic errors in the number density and other possible factors influencing the magnitude of $\beta$ \,\cite{Knoop2009ooa, Berninger2011uot}.

The dashed lines in Fig.\,\ref{comparison} show the results of the EFT fits for the two resonances.
For the high-field data in panel (a), we exclude from the fit the five data points with $a < 400\,a_0$, corresponding to the points below 557.6\,G in Fig.~\ref{ADold}(b). These points lie in the region where we suspect a strong influence by the $i/s$-wave crossing. Alternatively, we also include the lowest two data points in the fits (below the hatched region in Fig.~\ref{ADold}(b)), finding that this has negligible effect on the resulting value of $a_*$. For the low-field data set in panel (b) \cite{Knoop2009ooa}, the fit takes into account all data points.

Table\,\ref{fit} summarizes the fit parameters for the two atom-dimer resonances. The values $a_*$ obtained for the resonance positions are +653(25)\,$a_0$ and +419(10)\,$a_0$ for the high and low field features, respectively. The uncertainties include the statistical contributions and the uncertainties for the $a(B)$ conversion \cite{Berninger2013frw}. The difference in $a_*$ is remarkable and much larger than the uncertainties, while the values obtained for the width parameter $\eta_*$ are comparable within the error bars. The amplitude scaling factors resulting from the fit are $D=2.8(9)$ (high-field case) and $0.64(23)$ (low-field case), showing considerable deviations from unity with an opposite trend.

\subsection{Universal van der Waals theory}
\label{waals}

Recently, Wang and Julienne have introduced a new model \cite{Wang2013uvd} that builds in the pairwise van der Waals (vdW) interaction and, based on a numerical solution of the three-body Schr\"odinger equation, predicts the collision rate constants without any adjustable parameters. To describe the Feshbach resonance, the background scattering length $a_{\rm bg}$ and the resonance strength parameter $s_{\rm res}$ \cite{Chin2010fri} are needed as the two input parameters. For both cases considered here, we are in the regime of $s_{\rm res}\gg1$ (entrance-channel dominated resonances) and of a large $a_{\rm bg}/R_{\rm vdW}\gg1$. For Cs, $R_{\rm vdW}$ is equal to $101\,a_0$ \cite{Chin2010fri}.

The solid lines in Fig.\,\ref{comparison} show the results of the universal vdW model \cite{Wang2014pcs}. Although the theory, in principle, does not contain a free parameter, we introduce an additional amplitude scaling factor $D'$ to obtain an optimum fit with the experimental data; this is analogous to the parameter $D$ used in the EFT fit. The amplitude scaling factor takes into account possible amplitude variations between experiment and theory, which may result from various sources. On the experimental side, errors may result from the number density calibration and, on the theoretical side, the decay channels to deeply bound molecular states may not be properly taken into account because of the nonuniversal nature of these target states. Considerable amplitude deviations have been seen also in other experiments on atom-dimer resonances \cite{Bloom2013tou, Lompe2010ads, Nakajima2010nea}.

For the high-field resonance in Fig.\,\ref{comparison}(a), we find that the model describes its position and width very well, but an amplitude scaling factor of $D'=5.6$ is needed to fit the data (see Table~\ref{fit}). The predicted value $a_*$ for the loss maximum is $+625\,a_0$, which is consistent with the observed value $+653(25)\,a_0$ within $1.1\,\sigma$ of its uncertainty. For the low-field resonance in (b), the maximum appears at $+460\,a_0$, which is significantly (about $4\,\sigma$) above the experimental value $a_*=+419(10)\,a_0$. This deviation corresponds to 3\% of the Efimov period and may thus be considered as quite small. The required amplitude scaling factor $D' \approx 0.5$ is much smaller than in the high-field case.

\subsection{Discussion}
\label{Disc}

In the ideal Efimov scenario with its discrete scaling factor of $22.7$, the positions of all observables follow fixed ratios \cite{Braaten2006uif}. Those ratios, which are strictly valid only in the limit of very large $a$, provide benchmarks for testing the scenario in real atomic systems and for quantifying possible deviations. As such a benchmark, the position of the lowest atom-dimer resonance is ideally related to the one of the lowest triatomic resonance at $a_-$ by the ratio $a_*/|a_-| = 1.06$. With the experimentally determined values for $a_*$, as presented in Sec.\,\ref{EFT}, and the values for $a_-$ determined in our previous work \cite{Berninger2011uot}, we obtain $0.68(6)$ for the high-field region and $0.48(2)$ for the low-field region. These two experimental determinations of $a_*/|a_-|$ both lie substantially below the ideal value. This general trend is qualitatively expected based on theoretical approaches beyond the universal Efimov limit \cite{Dincao2009tsr, Platter2009rct, Thogersen2008upo}.

The two results for $a_*/|a_-|$ obtained for different Feshbach resonances deviate from each other, which points to the importance of the character of the underlying Feshbach resonance. Finite-range corrections as described to first order in terms of the effective range $k_0$ \cite{Chin2010fri} are not likely to explain the deviations, as $k_0$ shows only minor differences for both cases. This raises the question whether higher-order finite-range corrections may be relevant.

The universal vdW model, discussed in Sec.\,\ref{waals}, provides predictions remarkably close to the experimental observations and reproduces the central experimental findings that (i) the atom-dimer resonances are substantially down shifted as compared to the expectation from the universal Efimov limit, and that (ii) this down shift is smaller in the high-field region than in the low-field case. Comparing the two Feshbach resonances, the question arises whether the different values for $a_*/|a_-|$ can be mainly attributed to the difference in $s_{\rm res}$ or in $a_{\rm bg}$, or whether a combination of both is necessary to understand the situation.

An open issue concerns the amplitude of the observed atom-dimer resonances. In the high-field region the resonance amplitude is clearly larger than theoretically expected, which is quantified by the amplitude scaling parameters $D=2.8(9)$ and $D'=5.6(3)$ for the two fits applied. These values are too large to be explained by systematic experimental uncertainties, which we estimate to be below 50\%. For the previously observed atom-dimer resonance in the low-field region, the amplitude scaling factors $D=0.64(23)$ and $D'=0.48(7)$ are smaller than one, which indicates a trend opposite to the high-field region, but consistent with the observations of Refs.~\cite{Bloom2013tou, Lompe2010ads, Nakajima2010nea}. We can only speculate about possible causes for this difference. The collisional decay leads to more tightly bound molecular states and therefore involves coupling at short ranges. The present models apparently get the order of magnitude right, but they do not permit to describe the amplitude of the resonant decay on a fully quantitative level.

\section{Search for an atom-dimer avalanche effect}
\label{avalanche}

\begin{figure}[t]
\begin{centering}
\includegraphics[width=.50\textwidth] {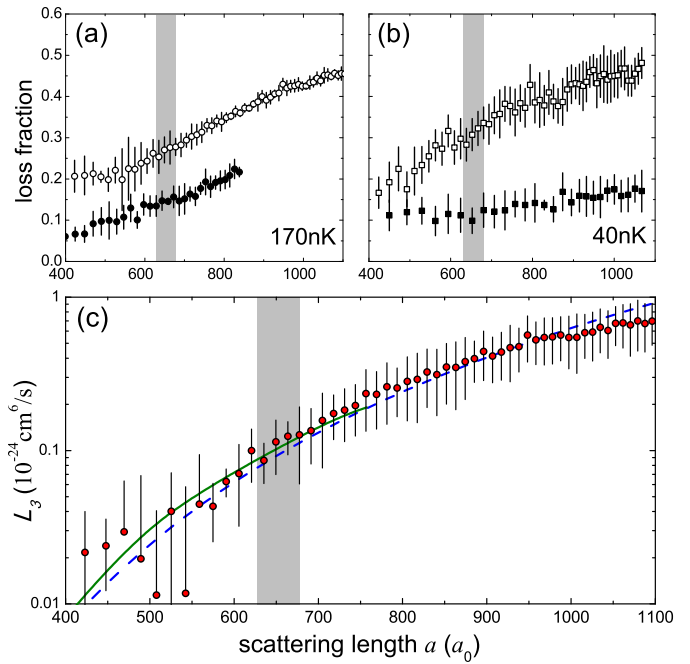}
\caption{(color online) Loss measurements in pure atomic samples. The loss fraction is measured with samples of (a) $1.5\times10^5$ atoms at a temperature of about 170\,nK and (b) $4\times10^4$ atoms at 40\,nK. In (a), the hold time is 2\,s for the open symbols and 400\,ms for the filled ones. In (b), the hold time is 1\,s for the open symbols and 50\,ms for the filled ones. (c) The recombination rate coefficient $L_3$ is extracted from the data set with a hold time of 2\,s from panel (a). The dashed line is the prediction from EFT \cite{Braaten2001tbr}, while the solid one is derived within the vdW model \cite{Wang2014pcs}. The error bars include the statistical uncertainties on the atom number. In all three panels, the grey region indicates the position of the loss resonance in atom-dimer mixtures; see Sec.\,\ref{discussion}. The width of this region reflects the uncertainty of the resonance's center position.}
\label{L3}
\end{centering}
\end{figure}

Three experimental groups have reported on the observation of atom-dimer resonances in measurements performed with purely atomic samples of $^{39}$K \cite{Zaccanti2009ooa} and $^7$Li \cite{Gross2009oou, Pollack2009uit}. These indirect observations have been attributed to an \textit{avalanche} process, during which the dimers formed in three-body recombination events collide elastically with the trapped atoms before leaving the sample. The energy released in a single recombination event is sufficient to kick several atoms out of the trap, which leads to enhanced losses. These measurements are still debated \cite{Machtey2012udi, Langmack2012tam, Langmack2013alr} as the atom-dimer peak position $a_*$ can only be inferred employing a collisional model. In this Section, we present measurements obtained in pure atomic samples. We show that they are well described by EFT and the vdW model, without any significant avalanche effect.

We have measured the fraction of lost atoms after a fixed hold time in a magnetic field range corresponding to $a$ between 400 and 1100\,$a_0$. The hold times have been chosen in order to have an observable loss fraction in the range between 10 and 50\%. First we performed our measurements with the atomic sample as described in Sec.\,\ref{Procedure}, having a temperature of 170\,nK and an initial peak number density of $1.6 \times 10^{12}$ cm$^{-3}$. Figure \ref{L3}(a) shows our results. As predicted, the loss fraction increases for larger values of $a$ as expected from the $a^4$ scaling \cite{weber2003tbr}, However, within our experimental uncertainties, the losses do not show any significant enhancements, neither at the atom-dimer resonance position $a_*$ nor at any other values of $a$. We have performed the loss measurements in samples with a higher peak density of $3.2 \times 10^{12}$ cm$^{-3}$, which are obtained in the course of a further evaporation step down to 40\,nK. Also with these experimental conditions we have observed no significant loss enhancement, as can be seen from the data Fig.\,\ref{L3}(b).

From the loss fraction data obtained at 170\,nK with a hold time of 2\,s, see open symbols in Fig.\,\ref{L3}(a), we extract the three-body recombination rate coefficient $L_3$. This is possible under the assumption that three-body collisions are the dominant loss mechanism and that heating is caused by the anti-evaporation effect \cite{weber2003tbr}. For values of $a$ below 500\,$a_0$, we observe additional background losses on a timescale exceeding tens of seconds. These background losses are subtracted in our data analysis. Figure \ref{L3}(c) shows our results on $L_3(a)$ together with the predictions of EFT \cite{note4} and the universal vdW model.

Our experimental data are consistent with the two loss models and this result suggests that losses in atomic samples of Cs under our experimental conditions can be predicted without including avalanche processes. Our results are consistent with earlier observations in the low magnetic field region that did not reveal any loss feature. Our observations are not consistent with predictions of a loss peak in Cs as discussed in \cite{Machtey2012udi}, but the model may not be appropriate for the specific situation of Cs \cite{Khaykovich2013pca}. The model of Ref.~\cite{Langmack2012tam} predicts a very broad feature of moderately enhanced losses near $a_*$. As it is experimentally very difficult to discriminate such a feature from the background, we cannot draw any conclusion on its presence.

A recent preprint reports on a search for the avalanche effect in heteronuclear atomic mixtures of $^{40}$K and $^{87}$Rb \cite{Ming2014sfa}. A narrow avalanche feature could not be observed, neither at values for $a$ where an atom-dimer resonance has been observed previously \cite{Zaccanti2009ooa, Gross2009oou, Pollack2009uit}, nor at other values. These observations are consistent with our findings and the suggested avalanche mechanism \cite{Zaccanti2009ooa} remains an unresolved issue.

\section{Conclusions}
\label{conclusion}

In conclusion, we have investigated inelastic atom-dimer collisions in mixtures of Cs atoms and Cs$_2$ dimers in the region of positive scattering lengths near the 550\,G Feshbach resonance. Our measurements reveal a resonance that results from the coupling of an atom and a dimer to an Efimov trimer state. We fit the data by using effective field theory predictions and we determine the resonance position and width. The resonance position $a^*=+653(25)a_0$ significantly deviates from the previous result obtained in the low magnetic field region, $a^*=+419(10)a_0$. For both resonances, their positions relative to the corresponding triatomic loss resonances strongly deviate from the ratio predicted for an ideal realization of Efimov's scenario in the large-$a$ limit. These observations demonstrate that universality is less robust in the positive-$a$ region than previously observed in the negative-$a$ region \cite{Berninger2011uot}, much more depending on the particular properties of the underlying Feshbach resonances used for interaction tuning.

We have compared our results with recent predictions obtained within the universal vdW model of Ref.\,\cite{Wang2013uvd}, which only requires the Feshbach resonance parameters and the vdW length to locate the Efimov features. The positions and the widths of the observed loss resonances are in very good agreement with the vdW model. For both resonances, the observed amplitudes differ strongly from the theoretical predictions. Surprisingly, they deviate in opposite directions for the two Feshbach resonance scenarios. Our results are an important step towards a complete understanding of Efimov processes in atomic systems. The extension of similar theoretical and experimental investigations to other species and to heteronuclear mixtures can probably shed new light on the appearance of Efimov states in real atomic systems, the robustness of universality, and on the influence of the particular Feshbach resonances on the Efimov spectrum.

Additional loss measurements carried out in purely atomic samples have not provided any signatures of an avalanche effect near the atom-dimer resonance position. The presence or absence of such features depending on the particular species is an unresolved issue that deserves more attention in future experiments.

\section*{Acknowledgements}
\label{Ack}

We thank P. S. Julienne and Y. Wang for discussions and for providing us with their model results, and J. D'Incao, L. Khaykovich, and E. Braaten for fruitful discussions. This work was supported by the Austrian Science Fund FWF within project P23106. A.~Z. was supported within the Marie Curie Project LatTriCs 254987 of the European Commission.


\begin{thebibliography}{56}
\bibitem[{not({\natexlab{a}})}]{note1}
\bibinfo{note}{For the present experiments, we have selected the 550-G Feshbach
  resonance instead of the even broader Feshbach resonance near 800\,G, because
  the latter does not provide an efficient entrance gate to populate the weakly
  bound molecular state.}
\bibitem[{not({\natexlab{c}})}]{note3}
\bibinfo{note}{We give the values for the energies in frequency units.}
\end{thebibliography}
\end{document}